\documentclass[aps,pra,twocolumn,amsmath,amssymb,showpacs]{revtex4}
\usepackage{amsmath}
\usepackage{graphicx}
\usepackage{epsfig}
\usepackage{amsfonts}

\begin{document}
\title{Cooling of a mirror in cavity optomechanics with a chirped pulse}
\author{Jie-Qiao Liao}
\affiliation{Department of Physics and Institute of Theoretical
Physics, The Chinese University of Hong Kong, Shatin, Hong Kong
Special Administrative Region, People's Republic of China}
\author{C. K. Law}
\affiliation{Department of Physics and Institute of Theoretical
Physics, The Chinese University of Hong Kong, Shatin, Hong Kong
Special Administrative Region, People's Republic of China}
\date{\today}

\begin{abstract}
We investigate the response of a harmonically confined mirror to an
optical pulse in cavity optomechanics. We show that when the pulsed
coupling strength takes the form of a chirped pulse, thermal
fluctuations of the mirror can be significantly transferred to the
cavity field. In addition, the frequency modulation of the pulse
could enable a better cooling performance by suppressing the
sensitivity of the dependence of detuning and pulse areas. Using
numerical investigations, we find that the pulsed-cooling is mainly
limited by the cavity-field decay rate.
\end{abstract}


\maketitle

\section{Introduction}

In cavity optomechanical
systems~\cite{KippenbergRev,MarquardtRev,KarraiRev,AspelmeyerRev},
the cooling of a mechanical resonator is important to study the
mechanical effects of light in the quantum
domain~\cite{Fabre1994,Tombesi1994,Heidmann1994,Bose1997,cat1,
Ian2008,Zoller2009,Eisert2009,Wallquist2010,Girvin2010,Liao2011,
Mancini2002,Ferreira2006,Vitali2007,Paternostro2007,Hartmann2008,Genes2008,Aspelmeyer2009}.
In addition, the reduction of thermal noise generally is a
requirement to implement various applications in quantum information
relying on optomechanical
couplings~\cite{Zhang2003,Stannigel2003,Tian2010,Paternostro2011}.
With the recent progress of cooling in cavity optomechanical systems
~\cite{Wilson-Rae2007,Marquardt2007,Gene2008,Li2011,Aspelmeyer2006,Heidmann2006,
Bouwmeester2006,Kippenberg2008,Schliesser2009,Wang2009}, it is
becoming possible to access quantum ground states. The cooling
techniques, such as the resolved sideband
cooling~\cite{Wilson-Rae2007,Marquardt2007,Gene2008,Kippenberg2008,Schliesser2009,Wang2009},
generally make use of a continuous light field that drives a
mechanical resonator into a steady state of lower temperature.
Recently, several authors have begun to explore the possibilities of
cooling and manipulating the states of mirrors with optical
pulses~\cite{Vanner2010,Cerrillo2011,Hofer2011,Fiore2011}.

Motivated by the fact that chirped pulses can lead to an efficient
population transfer in two-level
systems~\cite{Goswami2003,Shore2003,Berman1981,Zakrzewski1985,Hioe1984,Eberlybook},
we examine how a chirped-pulse interaction can transfer thermal
fluctuations from the mirror to the cavity field. As we shall
discuss below, the correspondence between a two-level system and
optomechanical systems [defined in Eq.~(\ref{Hamiltonian_S})] can be
established via the Heisenberg equations of motion of the linearized
system under a rotating wave approximation (RWA). Therefore,
analytical solutions known in two-level systems driven by chirped
pulses can be applied here~\cite{Eberlybook}. Owing to the frequency
modulation in chirped pulses, population transfer can be made
without the need for a high-precision control of detuning and pulse
areas.

In this paper, we will first provide a formulation of the linearized
quantum system driven by a general pulse. Then we will indicate how
the mirror-field coupling can be shaped into a chirped form by using
a proper time-dependence of an external driving field. Specifically,
we will study a class of chirped pulses proposed by Allen and Eberly
for two-level systems~\cite{Eberlybook}. Such a class of chirped
pulses has analytic solutions for nondissipative systems, and
depending on the chirped parameters, they describe both adiabatic
and nonadiabatic transitions. Using numerical calculations, we
include counter-rotating terms and dissipation effects, and we
demonstrate that cooling can be achieved with the chirped-pulse
coupling. In the case where the mechanical damping rate of the
mirror is sufficiently small, the cooling performance is mainly
limited by the cavity-field decay rate.

\section{Model}

\begin{figure}[ptb]
\center
\includegraphics[bb=58 657 270 757, width=3.3 in]{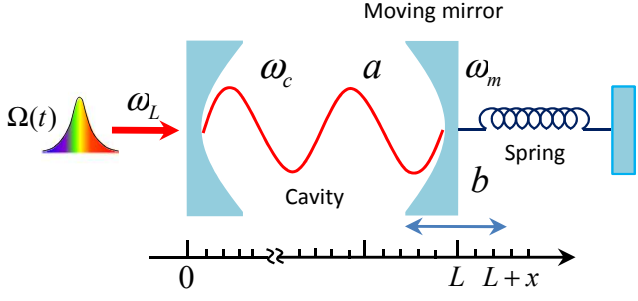}
\caption{(Color online) Schematic diagram of the cavity
optomechanical system. A Fabry-Perot cavity, formed by a fixed end
mirror and a harmonically bound end mirror, is driven by a
pulse.}\label{setup}
\end{figure}

Our model consists of a Fabry-Perot cavity formed by a fixed end
mirror and a moving end mirror connected with a spring
(Fig.~\ref{setup}). We consider a single cavity field mode with a
resonance frequency $\omega_{c}$ and creation (annihilation)
operator $a^{\dag}$ ($a$). The moving mirror is treated as a
quantum harmonic oscillator with a frequency $\omega_{m}$ and
creation (annihilation) operator $b^{\dag}$ ($b$).  Assuming the
cavity is driven by an external field with a carrier frequency
$\omega_{L}$ and a time-varying amplitude $\Omega(t)$, the
Hamiltonian of the system (in a rotating frame with the frequency
$\omega_{L}$) is given by,
\begin{eqnarray}
H_{S}&=&\hbar\Delta_{c}a^{\dag}a+\hbar\omega_{m}b^{\dagger}b-\hbar
ga^{\dag}a(b^{\dagger}+b)\notag\\
&&+\hbar\Omega(t)a^{\dag}+\hbar\Omega ^{\ast }(t)a,
\label{Hamiltonian_S}
\end{eqnarray}
where $\Delta_{c}=\omega_{c}-\omega_{L}$ is the detuning and $g$
is the radiation-pressure coupling strength.

To treat the damping and noise in our model, we consider the
system linearly coupled to oscillator baths. Under the Markovian
approximation and neglecting counter-rotating terms in system-bath
coupling, the quantum Langevin equations for the operators $a$ and
$b$ are given by,
\begin{subequations}
\label{Langevineq}
\begin{gather}
\dot{a}=-i\Delta _{c}a+iga(b^{\dagger }+b)-i\Omega(t)-\frac{\gamma _{c}}{2}a+a_{in},\\
\dot{b}=-i\omega _{m}b+iga^{\dag }a-\frac{\gamma _{m}}{2}b+b_{in},
\end{gather}
\end{subequations}
where $\gamma_{c}$ ($\gamma_{m}$) is the cavity field (mirror
motion) decay rate, $a_{in}$ is the vacuum radiation noise operator
for the cavity and $b_{in}$ is the mechanical noise operator for the
mirror. Both $a_{in}$ and $b_{in}$ have zero mean values and they
are characterized by the correlation functions $\langle
a_{in}(t)a^{\dagger}_{in}(t')\rangle=\gamma_{c}\delta(t-t')$,
$\langle a^{\dagger}_{in}(t)a_{in}(t')\rangle=0$, $\langle
b_{in}(t)b^{\dagger}_{in}(t')\rangle=\gamma_{m}(\bar{n}_{m}+1)\delta(t-t')$,
and $\langle
b^{\dagger}_{in}(t)b_{in}(t')\rangle=\gamma_{m}\bar{n}_{m}\delta(t-t')$,
where $\bar{n}_{m}=[\exp(\hbar\omega_{m}/k_{B}T_{m})-1]^{-1}$ is the
average thermal excitation number of the mirror at temperature
$T_{m}$. In this paper we will investigate the regime with $\omega_m
\gg \gamma_m$. This is a regime where the Markovian approximation
for the mirror noise can be justified \cite{Vitali2007}.

\section{Linearized system and formal solution}

By writing $o=\langle o\rangle+\delta o$ ($o=a,b$) and assuming the
fluctuations are small ($|\langle o\rangle|^2\gg \langle \delta
o^\dag \delta o \rangle$) during the pulse interaction, we may
linearize Eq.~(\ref{Langevineq}) and obtain the equations of motion,
\begin{subequations}
\label{fluctuateq}
\begin{gather}
\delta \dot{a}=-i\Delta(t)\delta a+ig\langle a\rangle(\delta
b^{\dagger}+\delta b)-\frac{\gamma _{c}}{2}\delta a+a_{in},\label{fluctuateqa}\\
\delta \dot{b}=-i\omega_{m}\delta b+ig[\langle
a^{\dagger}\rangle\delta a+\langle a\rangle\delta a^{\dag
}]-\frac{\gamma_{m}}{2}\delta b+b_{in}\label{fluctuateqb},
\end{gather}
\end{subequations}
with $\Delta(t)=\Delta_{c}-2g\textrm{Re}[\langle b(t)\rangle]$.
The $\langle a(t)\rangle$ and $\langle b(t)\rangle$ are governed
by:
\begin{subequations}
\label{clasequation}
\begin{gather}
\langle\dot{a}\rangle=-i\Delta(t)\langle
a\rangle-i\Omega(t)-\frac{\gamma_{c}}{2}\langle a\rangle,\\
\langle\dot{b}\rangle=-i\omega_{m}\langle b\rangle+ig|\langle
a\rangle|^{2}-\frac{\gamma_{m}}{2}\langle b\rangle.
\end{gather}
\end{subequations}
For the linear approximation made above, we have neglected
nonlinear terms $i g \delta a (\delta b+ \delta b^\dag)$ in
Eq.~(\ref{fluctuateqa}) and $i g \delta a^\dag \delta a$ in
Eq.~(\ref{fluctuateqb})~\cite{remark1}.

Equation~(\ref{fluctuateq}) corresponds to a linear coupling
described by the Hamiltonian $H_{I}=-\hbar g[\langle
a^{\dagger}(t)\rangle\delta a+\langle a(t)\rangle\delta
a^{\dagger}](\delta b^{\dagger}+\delta b)$. Here $\langle
a(t)\rangle$ modulates the mirror-field coupling and its time
dependence can be controlled by the driving amplitude $\Omega(t)$.
We point out that any desirable $\langle a(t)\rangle$ as a function
of time can in principle be achieved by a corresponding $\Omega(t)$,
according to Eq. (4a). For convenience, we let
\begin{equation}
g\langle
a^{\dagger}(t)\rangle\equiv\chi(t)e^{i\phi(t)}e^{-2ig\int_{0}^{t}\textrm{Re}[\langle
b(\tau)\rangle]d\tau}, \label{couplingstrength}
\end{equation}
where $\chi(t)$ and $\phi(t)$ are real functions, and the phase
angle $-2g\int_{0}^{t}\textrm{Re}[\langle b(\tau)\rangle]d\tau$ is
introduced in order to compensate for the phase shift induced by the
dynamical cavity frequency shift in $\Delta(t)$.

By defining operators $\delta A(t)=\delta a
e^{i\left[\phi(t)+\int_{0}^{t}\Delta(\tau)d\tau\right]}$ and $\delta
B(t)=\delta be^{i\omega_{m}t}$, Eq.~(\ref{fluctuateq}) can be
concisely written as $\mathbf{\dot{v}}(t)
=\mathbf{M}(t)\mathbf{v}(t)+\mathbf{N}(t)$, where
$\mathbf{v}(t)=[\delta A(t),\delta B(t),\delta A^{\dagger}(t),\delta
B^{\dagger}(t)]^{T}$, and
\begin{widetext}
\begin{eqnarray}
\mathbf{M}(t)=\left[\begin{array}{cccc}-\frac{\gamma_{c}}{2}
+i\dot{\phi}(t)&
i\chi(t)e^{i(\Delta_{c}-\omega_{m})t}& 0 & i\chi(t)e^{i(\Delta_{c}+\omega_{m})t}\\
i\chi(t)e^{-i(\Delta_{c}-\omega_{m})t}& -\frac{\gamma_{m}}{2} &i\chi(t)e^{i(\Delta_{c}+\omega_{m})t}&0 \\
0 &-i\chi (t)e^{-i(\Delta_{c}+\omega_{m})t} & -\frac{
\gamma _{c}}{2}-i\dot{\phi}(t)& -i\chi(t)e^{-i(\Delta_{c}-\omega_{m})t}\\
-i\chi(t)e^{-i(\Delta_{c}+\omega_{m})t} & 0 &
-i\chi(t)e^{i(\Delta_{c}-\omega_{m})t}& - \frac{\gamma _{m}}{2}
\end{array}
\right],\label{mMatrix}
\end{eqnarray}
\end{widetext}
and $\mathbf{N}(t)=[A_{in}(t),B_{in}(t),
A_{in}^{\dagger}(t),B_{in}^{\dagger}(t)]^{T}$ with
$A_{in}(t)=a_{in}e^{i\left[\phi(t)+\int_{0}^{t}\Delta(\tau)d\tau\right]}$
and $B_{in}(t)=b_{in}e^{i\omega_{m}t}$. The solution of
$\mathbf{v}(t)$ is
\begin{equation}
\mathbf{v}(t)=\mathbf{G}(t)\mathbf{v}(0)+\mathbf{G}(t)
\int_{0}^{t}\mathbf{G}^{-1}(\tau)\mathbf{N}(\tau)d\tau,
\end{equation}
where $\mathbf{G}(t)$ is governed by
\begin{equation}
\mathbf{\dot{G}}(t)=\mathbf{M}(t)\mathbf{G}(t),\label{eq8}
\end{equation}
with $\mathbf{G}(0)=I$ being the identity matrix.

The state of the system can be conveniently described by  a
covariance matrix $\mathbf{R}(t)$ whose elements are:
$\mathbf{R}_{ll'}(t)=\langle
\mathbf{v}_{l}(t)\mathbf{v}_{l'}(t)\rangle$ ($l,l'=1,2,3,4$).
Therefore $\mathbf{R}_{31}(t)=\langle\delta a^{\dagger }\delta
a\rangle$ and $ \mathbf{R}_{42}(t) = \langle \delta b^{\dagger
}\delta b\rangle$ are mean {\em displaced particle numbers}
measuring the fluctuations. By Eq. (7), $\mathbf{R}(t)$ reads,
\begin{equation}
\mathbf{R}(t)=\mathbf{G}(t)\mathbf{R}(0)\mathbf{G}^{T}(t)+\mathbf{G}(t)\mathbf{Z}
(t)\mathbf{G}^{T}(t),\label{expreofR}
\end{equation}
with
\begin{equation}
\mathbf{Z}(t)=\int_{0}^{t}\int_{0}^{t}\mathbf{G}
^{-1}(\tau)\mathbf{C}(\tau,\tau^{\prime})[\mathbf{G}^{-1}( \tau
^{\prime})]^{T}d\tau d\tau ^{\prime}.
\end{equation}
The matrix $\mathbf{R}(0)$ is determined by the initial condition of
the system, and $\mathbf{C}(\tau,\tau^{\prime})$ is the two-time
correlation function of noise operators, which is defined by the
elements $\mathbf{C}_{l,l^{\prime }}(\tau,\tau^{\prime})=\langle
\mathbf{N}_{l}(\tau)\mathbf{N}_{l^{\prime }}(\tau^{\prime})\rangle$
($l,l'=1,2,3,4$). Assuming that initially the cavity is in vacuum
and the mirror is in a thermal equilibrium at the same temperature
$T_{m}$ as its bath, i.e., $\rho(0)=|0\rangle_{c}\langle
0|_{c}\otimes\rho_{th}(T_{m})$ with $
\rho_{th}(T_{m})=\exp[-\hbar\omega_{m}b^{\dagger}b/k_{B}T_{m}]/\textrm{Tr}(\exp[-\hbar\omega_{m}b^{\dagger}b/k_{B}T_{m}])$,
then the matrix $\mathbf{R}(0)$ has three nonzero elements:
$\mathbf{R}_{13}(0)=1$, $\mathbf{R}_{24}(0)=\bar{n}_{m}+1$, and
$\mathbf{R}_{42}(0)=\bar{n}_{m}$. In addition, the Markovian baths
imply $\mathbf{C}(\tau,\tau')=\mathbf{C}\delta(\tau-\tau^{\prime})$,
where $\mathbf{C}$ is a constant matrix with three nonzero elements:
$\mathbf{C}_{13}=\gamma_{c}$,
$\mathbf{C}_{24}=\gamma_{m}(\bar{n}_{m}+1)$, and
$\mathbf{C}_{42}=\gamma_{m}\bar{n}_{m}$.

\section{Cooling of the Mirror}

We will employ the expectation value of displaced phonon number
$\langle \delta b^{\dagger }\delta b\rangle$ as an indicator of
cooling. This means that the idea of cooling in our scheme should
be understood as a process of reducing excitations with respect to
the mean amplitude $\langle b \rangle$ of the mirror.

\subsection{Chirped-pulse coupling}

We now ask what $\Omega(t)$ is suitable for cooling the mirror.
Guided by the fact that a chirped-pulse driving can efficiently
realize population transfer in two-level
systems~\cite{Goswami2003,Shore2003,Berman1981,Zakrzewski1985,Hioe1984,Eberlybook},
we consider the coupling strength in Eq.~(\ref{couplingstrength})
taking the chirped form \cite{Eberlybook},
\begin{subequations}
\label{chirpedpulse}
\begin{gather}
\chi(t)=\chi_{0}\textrm{sech}[\alpha(t-t_{0})],\label{chirpedpulsesech}\\
\theta(t)=\dot{\phi}(t)=\beta \tanh[\alpha(t-t_{0})].
\end{gather}
\end{subequations}
Here $t_0$ determines the time of the pulse peak entering the
cavity, $\alpha^{-1}$ measures the pulse duration, $\beta$ controls
the magnitude of the frequency modulation, and $\chi_0$ is the
strength of the pulse coupling.

\begin{figure}[ptb]
\center
\includegraphics[bb=24 25 432 229, width=3.3 in]{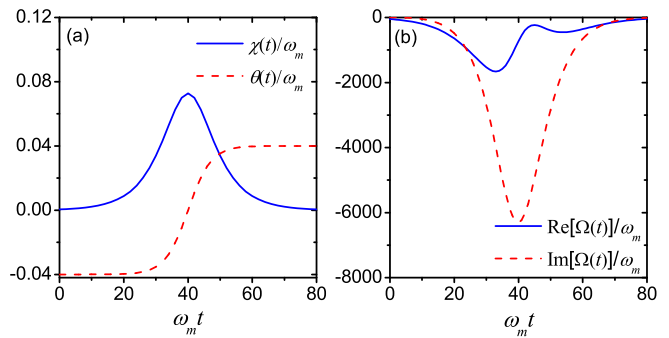}
\caption{(Color online) Plot of (a) the chirped pulse given in
Eq.~(\ref{chirpedpulse}) and (b) the original driving amplitude
$\Omega(t)$ vs the scaled time $\omega_{m}t$.  The parameters are:
$\alpha/\omega_{m}=0.14$, $\beta/\omega_{m}=0.04$,
$\omega_{m}t_{0}=40$, and
$\chi_{0}=\frac{1}{2}\sqrt{\alpha^{2}+\beta^{2}}$.
}\label{chirpedpulsefig}
\end{figure}

The required driving amplitude $\Omega(t)$ for generating the above
$\chi(t)$ and $\phi(t)$ can be found using
Eqs.~(\ref{clasequation}), (\ref{couplingstrength}), and
(\ref{chirpedpulse}) as
\begin{equation}
\Omega(t)=i\langle
\dot{a}(t)\rangle-\left[\Delta(t)-i\frac{\gamma_{c}}{2}\right]\langle
a(t)\rangle,
\end{equation}
with
\begin{subequations}
\label{clasmotion}
\begin{gather}
\langle
a(t)\rangle=\frac{\chi_{0}}{g}\text{sech}[\alpha(t-t_{0})]e^{-i\left[\phi(t)-2g\int_{0}^{t}\textrm{Re}[\langle
b(\tau)\rangle]d\tau\right]},\label{av-a}\\
\langle
b(t)\rangle=i\frac{\chi_{0}^{2}}{g}\int_{0}^{t}\text{sech}^{2}[\alpha(\tau-t_{0})]e
^{-\left(\frac{\gamma_{m}}{2}+i\omega_{m}\right)(t-\tau)}d\tau,\label{av-b}\\
\phi(t)=\frac{\beta}{\alpha}\log\left[\frac{\cosh[\alpha(t-t_{0})]}{\cosh(\alpha
t_{0})}\right].
\end{gather}
\end{subequations}
An example illustrating the chirped pulse coupling and the
corresponding $\Omega(t)$ is given in Fig.~\ref{chirpedpulsefig}.

\subsection{Ideal case}

To find the optimal relation for $\chi_{0}$, $\alpha$, and $\beta$
in Eq.~(\ref{chirpedpulse}) such that efficient cooling can be
achieved, we first investigate the nondissipative case
($\gamma_{c}=0$ and $\gamma_{m}=0$) as a guide. Specifically, we
consider the resonance case $\Delta_{c}=\omega_{m}$. Under the
condition $(\Delta_{c}+\omega_{m})\gg\chi_{0}$, we discard the terms
$\pm i\chi(t) e^{\pm i(\Delta_{c}+\omega_{m})t}$ in
Eq.~(\ref{mMatrix}) by RWA, then by letting $u(t)=\langle \delta
A^{\dagger}\delta B\rangle+\langle\delta B^{\dagger}\delta
A\rangle$, $v(t)=i(\langle \delta B^{\dagger}\delta
A\rangle-\langle\delta A^{\dagger}\delta B\rangle)$, and
$w(t)=\langle \delta A^{\dagger}\delta A\rangle-\langle\delta
B^{\dagger}\delta B\rangle$, we can obtain the Bloch equations
\begin{subequations}
\label{Blocheq}
\begin{gather}
\dot{u}(t)=\dot{\phi}(t)v(t),\\
\dot{v}(t)=-\dot{\phi}(t)u(t)+2\chi(t)w(t),\\
\dot{w}(t)=-2\chi(t)v(t).
\end{gather}
\end{subequations}
Under the initial condition $u_{i}=0$, $v_{i}=0$, and
$w_{i}=-\bar{n}_{m}$, we find that, when
\begin{equation}
\chi_{0}=\frac{1}{2}\sqrt{\alpha^{2}+\beta^{2}},\label{optirelation}
\end{equation}
the solution of the Bloch equations is~\cite{Eberlybook}
\begin{subequations}
\label{Blocheqsol}
\begin{gather}
u(t)=-\frac{\beta}{\alpha}v(t)=\frac{\bar{n}_{m}\beta}{2\chi_{0}}
\textrm{sech}[\alpha(t-t_{0})],\\
w(t)=\bar{n}_{m}\tanh[\alpha(t-t_{0})].
\end{gather}
\end{subequations}
Therefore the average quasi-phonon number evolves as
\begin{equation}
\langle \delta b^{\dagger}\delta b\rangle=\langle \delta
B^{\dagger}\delta
B\rangle=\frac{\bar{n}_{m}}{2}(1-\tanh[\alpha(t-t_{0})]).\label{phononnum}
\end{equation}
When $\alpha(t-t_{0})\gg1$, we have $\langle \delta
b^{\dagger}\delta b\rangle\approx0$ (for example,
$\tanh5=0.999909$), which implies that thermal noise in the mirror
can be extracted almost completely. Thus,
$\chi_{0}=\frac{1}{2}\sqrt{\alpha^{2}+\beta^{2}}$ is a relation to
implement efficient cooling of the mirror in the absence of
dissipation.

It is useful to note that for the constant-pulse coupling case
[i.e., $\chi(t)$ is a constant], one can also extract energy from
the mirror, but the corresponding solution is oscillatory at a Rabi
frequency, i.e., the phonon number in the mirror is a cosine
function of time. In this case the timing of the constant pulse is
crucial in order to locate the instant when the phonon number is
minimum. However, our scheme does not have such a timing control
issue because the solution~(\ref{phononnum}) indicates that after
the pulse duration (about $2t_{0}$), the residual quasiphonon number
of the mirror changes slowly in time by the behavior of the
$\tanh[\alpha(t-t_{0})]$ function.

As a remark, we indicate that $\chi_{0}$ should be much smaller than
the mirror frequency $\omega_{m}$ in order to meet the condition of
RWA. The violation of RWA would mean that the parametric interaction
of the form $\delta a^\dag \delta b ^\dag + h.c.$ becomes important,
which generally leads to heating of the system. Therefore by
Eq.~(\ref{optirelation}), the pulse duration characterized by
$\alpha^{-1}$ cannot be arbitrarily short, i.e.,
$\alpha<2\chi_{0}\ll 4\omega_{m}$. Our numerical calculations
(without RWA) indicate that the parameters used in
Fig.~\ref{phononphotonf} are quite sufficient for RWA to be valid.

\subsection{Dissipative case}

\begin{figure}[ptb]
\center
\includegraphics[bb=24 25 432 229, width=3.3 in]{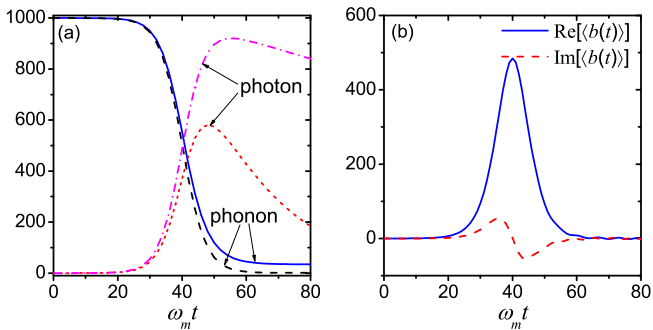}
\caption{(Color online) (a) Time evolution of the displaced phonon
number $\langle\delta b^{\dagger}\delta b\rangle$ and displaced
photon number $\langle\delta a^{\dagger}\delta a\rangle$ in the
chirped-pulse coupling case for two different values for the
$\gamma_{c}$. The solid and short dashed curves are for the case of
$\gamma_{c}/\omega_{m}=0.0435$, while the dashed and dash-dotted
curves are for the case of $\gamma_{c}/\omega_{m}=0.00435$. (b) Plot
of the real and imaginal parts of $\langle b(t)\rangle$ vs the
scaled time $\omega_{m}t$. Here
$\Delta_{c}=\omega_{m}$.}\label{phononphotonf}
\end{figure}

In realistic experiments, the interactions with environments will
inevitably lead to dissipation of the system. In addition, the
counter-rotating terms $(\delta a^\dag \delta b ^\dag + h.c.)$
ignored in RWA may modify the dynamic process. In what follows, we
numerically study the cooling process in the dissipative case beyond
RWA. We consider the realistically experimental parameters for the
system~\cite{Kippenberg2008}: $\omega_{m}\approx2\pi\times73.5$~MHz,
$\gamma_{m}\approx2\pi\times1.3$~kHz,
$\gamma_{c}\approx2\pi\times3.2$~MHz, and
$g\approx2\pi\times843.1$~Hz. Namely,
$\gamma_{m}/\omega_{m}\approx1.768\times10^{-5}$,
$\gamma_{c}/\omega_{m}\approx0.0435$, and
$g/\omega_{m}\approx1.147\times10^{-5}$. With these parameters and
the chirped pulse given in Fig.~\ref{chirpedpulsefig}, we solve
Eq.~(\ref{expreofR}) numerically. In Fig.~\ref{phononphotonf}(a), we
plot the time evolution of the mean displaced particle numbers. We
see that the $\langle\delta b^{\dagger}\delta b\rangle$ decreases
rapidly from its initial value ($\bar{n}_{m}=1000$) to a relatively
small number (about $34$) when the chirped pulse is applied. At the
same time, $\langle\delta a^{\dagger}\delta a\rangle$ in the cavity
increases rapidly from zero to a peak value and then decreases to
zero gradually through the cavity decay channel. We note that the
residual fluctuations of the mirror are limited by the noise of the
system, mainly of the cavity-field damping. Our numerical
investigations show that when the cavity-field decay rate is
$\gamma_{c}/\omega_{m}\approx0.00435$, the residual $\langle \delta
b^{\dagger }\delta b\rangle$ can further be reduced to $1.08$ [the
dash line in Fig.~\ref{phononphotonf}(a)]. In addition, numerical
calculations indicate the correction from the counter-rotating terms
is negligible with these parameters.

We remark that during the pulse interaction, the mirror attains a
non-zero coherent amplitude (i.e., $\langle b(t) \rangle \ne 0$)
according to Eq.~(\ref{av-b}). But such a coherent motion should not
be confused with the fluctuations we aim to reduce in this paper.
For the present parameters used in Fig.~\ref{phononphotonf}(a), we
plot the time evolution of $\langle b(t) \rangle$ in
Fig.~\ref{phononphotonf}(b). We see that there is a small amplitude
$|\langle b(2t_{0})\rangle|\approx 2$ near the end of the pulse
interaction. Actually, when the thermal fluctuation of the mirror is
completely transferred to the cavity, the mirror in the displaced
representation will be in its ground state. Therefore, in the
original representation, the mirror is prepared in a coherent state
of the mechanical motion.

We also point out that although there are residual cavity photons
after the pulse duration (around $2t_0$), the heating due to such
photons is found to be negligible because of the weak coupling
strength $g$. We can estimate that for a residual cavity photon
number $n_r$ at time $2t_0$, the mirror can be excited to have
phonon number $(gn_r/\omega_m)^2$, which is of the order of $0.001$
with the parameters used in Fig.~\ref{phononphotonf}
$(\gamma_{c}/\omega_{m}=0.0435)$. Therefore the heating is mainly
due to the heat bath of the mirror. For example, we find that the
heat bath of the mirror would increase the phonon number from $34$
to $38$ when the time evolves from $\omega_m t=80$ to $300$.

\subsection{Effects of the $\beta$ parameter}

\begin{figure}[ptb]
\center
\includegraphics[bb=15 22 430 231, width=3.3 in]{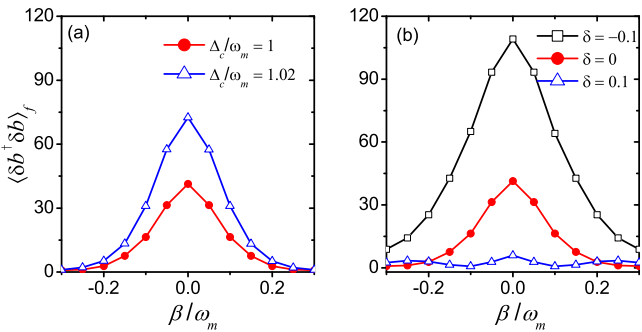}
\caption{(Color online) (a) Plot of the final mean displaced phonon
number $\langle\delta b^{\dagger}\delta b\rangle_f$ vs the phase
modulation amplitude $\beta$ for various detunings
$\Delta_{c}/\omega_{m}=1$ and $1.02$. (b) Plot of $\langle\delta
b^{\dagger}\delta b\rangle_f$ vs $\beta$ for various parameters
$\delta=-0.1$, $0$, and $0.1$.}\label{betaeffect}
\end{figure}

The frequency modulation characterized by the parameter $\beta$ is a
main feature of the chirped coupling. In the case of $\beta=0$, the
coupling corresponds to a $\pi$ pulse because
$\int_{-\infty}^{\infty}\alpha\textrm{sech}[\alpha(t-t_{0})]dt=\pi$
is the pulse area. However, such a simple $\pi$ pulse generally does
not bring an optimal cooling when dissipation and counter-rotating
terms are included. The parameter $\beta$ therefore provides a way
to adjust the pulse for a better cooling performance. In
Fig.~\ref{betaeffect}, we demonstrate this feature numerically by
plotting the final mean displaced phonon number of the mirror
(defined at time $t=2t_0$) as a function of $\beta$. There are two
situations [Fig.~\ref{betaeffect}(a) and \ref{betaeffect}(b)] that
we will discuss below, but in both figures, it is apparent that
non-zero values of $\beta$ can better reduce the displaced phonon
number of the mirror.

We point out that the final displaced phonon number can become
less sensitive to the detuning $\Delta_c$ when $|\beta|$ is
increased. This is shown in Fig.~\ref{betaeffect}(a) where we can
compare the sideband resonance case $\Delta_c = \omega_{m}$ with a
slightly off resonance case. We see that although cooling with an
off-resonance $\Delta_c$ is less effective, the dependence on
$\Delta_c$ becomes weaker as $|\beta|$ increases. For the
parameters used in Fig.~\ref{betaeffect}(a), the final displaced
phonon numbers are essentially the same when $|\beta| > 0.2
\omega_m$. This shares a similar feature in two-level systems as a
chirped pulse can make efficient population transfer in the
presence of inhomogeneous broadening.

In Fig.~\ref{betaeffect}(b), we illustrate the effect of $\beta$
on cooling when there are uncertainties in controlling the pulse
area. Such an error, for example, may come from an inaccurate
value of the coupling strength $g$. Let us express $\chi(t)$ as
\begin{equation}
\chi(t)=(1+\delta)\chi_{0}\textrm{sech}[\alpha(t-t_{0})],
\end{equation}
with $\delta$ describing the deviation. At $\beta=0$,
Fig.~\ref{betaeffect}(b) shows that a modest change of $\delta$ can
affect the final displaced phonon number quite significantly. In
fact, we notice that the case $\delta=0.1$ in the figure actually
corresponds to a better cooling. This indicates that $\chi_0$ in
Eq.~(\ref{optirelation}) is no longer optimal for cooling because of
dissipative effects. The search for optimal pulse parameters relies
on numerical work, but Fig.~\ref{betaeffect}(b) suggests that the
frequency modulation with a suitable range of $\beta$ may ease the
sensitivity of $\delta$ and hence improve the cooling performance
even though the pulse parameters are not exactly optimal.

\section{Conclusion and remarks}

To conclude, we have proposed a method to cool a moving mirror in
cavity optomechanics by a chirped pulse. Within the linearization
framework, we have shown how a chirped pulse coupling can be
achieved by an external driving field, and numerically we have
demonstrated that thermal fluctuations in the mirror can be
significantly transferred to the cavity after the pulse.  In
particular, the frequency modulation plays a positive role in the
cooling process especially when there are uncertainties in
controlling the detuning and pulse areas.

Finally, we remark that it would be difficult to present a general
comparison of the cooling efficiency between our scheme and the
resolved sideband cooling. This is because Eq.~(\ref{eq8}) has no
analytic solution and so the residual phonon number can only be
calculated numerically. Nevertheless, we notice that by decreasing
the cavity decay rate, the residual phonon number can be lowered. As
a specific example, with $\gamma_c/\omega_m=0.001$ and the same
other parameters as in Fig.~\ref{phononphotonf}, the residual phonon
number can reach $0.64$. Therefore, the system under such parameters
may effectively be considered as the ground state, although this
residual phonon number is higher than the resolved sideband cooling
limit $(\gamma_c/4\omega_m)^2$~\cite{Marquardt2007}. The main
purpose of this paper is to provide an alternative method of cooling
based on pulsed interaction, which is a transient solution rather
than a steady-state one. In other words, the process can occur in a
finite duration of time, and this could be a useful feature for
manipulating quantum states of the mirror.

\begin{acknowledgments}
This work is partially
supported by a grant from the Research Grants Council of Hong Kong,
Special Administrative Region of China (Project No.~CUHK401810).
\end{acknowledgments}


\begin{thebibliography}{99}
\bibitem{remark1}           For the pulsed coupling discussed in Sec. IV, we have numerically checked that $|\langle a\rangle|^{2}\gg
                            \langle \delta a^{\dagger}\delta a\rangle$ and $ |\langle b\rangle|^{2}
                            \gg \langle \delta b^{\dagger}\delta b\rangle$ during the pulse duration
                            ($0<t<2t_{0}$).
\end{thebibliography}
\end{document}